\documentclass[12pt,preprint]{aastex}

\begin{document}

\title{A Magnetic Bomb Scenario for Relativistic Jet Events in the Microquasar GRS 1915+105}

\author{Stephen S. Eikenberry\altaffilmark{1} \& Maurice H.H.M. van Putten\altaffilmark{2}}

\altaffiltext{1}{Astronomy Department, University of Florida, Gainesville, FL  32611}

\altaffiltext{2}{LIGO Project, NW17-161, 175 Albany Street, Cambridge, MA 02139-4307, USA}

\begin{abstract}

	We present a magnetic bomb scenario for the multiwavelength
behavior during ``Type B" relativistic jet events in the microquasar
GRS 1915+105. These events are characterized by a hard X-ray dip which
terminates in a soft X-ray spike.  The scenario, based on the
suspended accretion model for long gamma-ray bursts, posits a magnetic
dynamo around an accreting Kerr black hole which reaches the Van
Putten-Levinson critical value of poloidal magnetic field
energy-to-kinetic energy (${\cal E}_B/{\cal E}_k \simeq 1/15$) in the
inner accretion disk. The toroidal inner accretion disk subsequently
becomes unstable, and the poloidal magnetic field energy in the inner
torus magnetosphere is promptly released as it disconnects from the
black hole -- the magnetic bomb (``B-bomb'') explosion.  This scenario
matches the long-duration time-scale and spectral evolution of the
hard X-ray dip, and the short-duration time-scale and energetics of
the soft X-ray spike of type B events, as well as correlated features
in the infrared and radio wavebands.  We discuss some implications of
these results for understanding the formation of relativistic jets in
black hole systems.

\end{abstract}

\keywords{black hole physics -- infrared: stars -- stars: individual (GRS 1915+105) -- X-rays: stars}

\section{Introduction}

     Since the initial discovery of their relativistic jets
(\citet{MR94} ; \citet{Tingay} ; \citet{HR}), the Galactic
microquasars have shown great potential as a laboratory for studying
the formation of relativistic jets in black hole systems.  These
systems contain black holes of a few stellar masses, i.e.: $M_{BH}
\simeq 7 M_{\odot}$ for GRO 1655-40 \citep{Orosz} and $M_{BH} \simeq
14 M_{\odot}$ for GRS 1915+105 \citep{Greiner}, as opposed to the
supermassive black holes of $\sim 10^8-10^9 M_{\odot}$ in quasars and
active galactic nuclei.  As a consequence, the timescales for
variability are typically much shorter in microquasars, ranging from
milliseconds to hours, as opposed to years to centuries in quasars and
AGN.  This allows studies of repeated patterns of jet-producing
behaviors on timescales feasible for human investigation.

	GRS 1915+105 is perhaps the most promising observational
target for understanding the microquasar phenomenon, due to its
continued violent activity in X-ray, infrared, and radio wavebands.
\citet{MR94} first discovered relativistic jets in GRS 1915+105
through observations of oppositely-directed, collimated radio blobs
moving with $v \simeq 0.92c$.  Several years later, \citet{FP97}
observed fast infrared (IR) flares and, later the same day, radio
flares, hypothesized to be smaller-scale versions of the ejections
seen by Mirabel \& Rodriguez.  Subsequent multiwavelength observations
revealed a rich phenomenology of X-ray flares and dips, IR flares, and
radio flares. This confirmed that GRS 1915+105 provided the first
time-resolved observations of relativistic jet formation
(\citet{Eiken98a}; \citet{Eiken98b} ; \citet{Felix98} ; \citet{FP99}).

	GRS 1915+105 shows (at least) three types of jet-forming
activities \citep{Eiken01}.  The first of these, Type A, produces the
major ejection events \citep{MR94}, and is associated with a reduction
in hard X-ray flux on long timescales \citep{Harmon97}.  The second of
these types, Type B, shows smaller radio/infrared flares associated
with X-ray dips/spikes (\citet{Eiken98a}; \citet{Felix98}), and also
exhibits apparent superluminal motions \citep{Vivek}.  The third type,
Type C, shows much fainter IR flares and associated radio flares with
soft X-ray dips.

	We present here a magnetic bomb scenario as a new mechanism
for the multiwavelength behavior of Type B events.  Our scenario is
based on the suspended accretion model for long gamma-ray bursts
\citep{mvp01} and the van Putten-Levinson stability criterion for
magnetized disks \citep{mvp03}. This posits a rapidly spinning Kerr
black hole \citep{ker63} surrounded by a uniformly magnetized inner
accretion disk, represented by {\em two counter-oriented current
rings} which support a inner toroidal magnetosphere around the black
hole, as well as an outer toroidal magnetosphere which extends to
infinity. Prompt disconnection of the inner torus magnetosphere from
the black hole releases magnetic energy in the form of radiation -- a
magnetic explosion (``B-bomb"). We shall quantify the energies and
timescales of this mechanism, as well as the preceeding evolution of
build-up of the torus and its magnetosphere.  This analysis describes
a proposed two-state three-way ``hole-torus-jet" connection in the
low-hard dip state of Type B events. We note also that as we finished
preparing this work for submission, \citet{liv03a} have independently
announced some very interesting, related work on a two-state
``disk-jet" connection in GRS 1915+105.

The interaction of the black hole and the surrounding uniformly
magnetized torus operates similarly to pulsar magnetospheres when
viewed in poloidal topology.  The inner face of the torus facing the
black hole event horizon is equivalent to a pulsar, once we identify
the event horizon of the black hole with a ``compactified infinity''
endowed with non-zero angular velocity \citep{mvp99}.  The geometry of
this black hole and torus system viewed in poloidal cross-section shows
a leading-order effect in the magnetic coupling (Figure 1), so that
most of the black hole event horizon surface provides the asympotic
end point for the open magnetic field lines of the inner torus
magnetosphere. A rapidly spinning black hole will rotate much faster
than the inner accretion disk for nominal values of the disk
radius. By the equivalence to pulsar magnetospheres described above,
the torus and the black hole interact, whereby (a) most of the black
hole spin energy is dissipated in the event horizon and (b) most of
the black hole luminosity is incident onto the inner face of the torus
(Figure 2).

In the weak magnetic field limit, extraction of energy from a Kerr
black hole by a torque to surrounding matter in the accreting state is
described by \citet{ruf75}.  We appeal to the above equivalence to
pulsars for the extension to the strong magnetic field-limit. This
torque creates a ``magnetic wall" around the black hole \citep{mvp03}
blocking passage of accreting material through the inner disk -- a
``suspended accretion'' state \citep{mvp01a} develops a torus as
matter continues to build up by accretion in the extended disk. The
black hole gradually spins down by conservation of energy and angular
momentum. 

Note that the equivalence to pulsar magnetospheres establishes
causality in the extraction of black hole-spin energy by horizon
Maxwell stresses, circumventing the criticism of \citet{pun90} to the
force-free limit of \citet{bla77} -- see \citet{mvp03}. Also, the
field-lines connecting the black hole with the disk are equivalent to
those on the outer face of the torus which extend to infinity and are open, 
not closed as claimed wrongly by \citep{li00}. Closed field-lines in the
inner and outer torus magnetosphere each form a toroidal 'bag' \citep{mvp99}.

In \S2 below, we describe the multiwavelength phenomenology of Type B
events.  In \S3, we present the overall picture for the magnetic bomb
scenario, based on the suspended accretion model for long gamma-ray
bursts \citep{mvp01}. In \S4, we identify the time-scales, energetics
and spectral evolution of the X-ray lightcurves in transition through
a critical point of stability of the inner disk surrounding the black
hole. In \S5, we discuss the implications of this scenario for
relativistic jet formation in GRS 1915+105 and future prospects for
investigating its validity.  Finally, in \S6, we present our
conclusions.

\section{The Overall Multi-wavelength Pattern of Type B Events}

        \citet{Eiken98a} observed GRS 1915+105 on August 13-15, 1997
using the Proportional Counter Array (PCA) instrument of the Rossi
X-ray Timing Explorer (RXTE), as well as an infrared camera mounted on
the Palomar 200-inch telescope.  As reported in \citet{Eiken98a} and
\citet{Eiken98b}, on August 14-15 GRS 1915+105 exhibited a complex,
but repeating, pattern of X-ray activity associated with IR flares.
The pattern begins with the onset of a spectrally-hard X-ray dip,
shown in Figure 3, where the emission is dominated by a non-thermal
power-law spectral component.  This is the so-called ``$\beta$'' class
of variability described by \citet{Tomasso2}.  After a duration of
$T_{dip} \sim 600$s, the end of the dip is marked by a spectrally-soft
X-ray spike, followed by a recovery to a higher count rate, which is
dominated by thermal emission from the accretion disk (see
e.g. \citet{Tomasso1}). The X-ray lightcurve then transitions to an
oscillating state characterized by rapid changes in the blackbody
spectral component coupled with soft flares in the power-law component
\citep{Eiken00}.  Finally, a new hard dip begins.

	Figures 4 and 5 show X-ray spectroscopic parameters for GRS
1915+105 during this time period.  We extracted X-ray spectra in the
$\sim 3-25$ keV range at a time resolution of 1 second.  We applied
standard procedures for response matrix generation for background
estimation and subtraction, and corrected for PCA deadtime.  We used
the XSPEC v.11.2 software package to fit each spectrum with a standard
model for black hole candidates consisting of a ``soft'' component
modeled as a multi-temperature disk black body
(e.g. \citet{Mitsuda84}) and a ``hard'' component modeled as a
power-law, plus hydrogen absorption fixed to a column density of $6
\times 10^{22} \ {\rm cm^{-2}}$ \citep{Muno99}.  We added a systematic
error of 1\% to each spectrum before the fit was performed.  We
obtained typical reduced chi-squared values $\chi^2_{\nu} \simeq 1.5$.
(We present more detailed discussion of this behavior in Section 4.1.)

	The lightcurve of the dip-terminating X-ray spike itself has
several important features.  The X-ray count rates rise to maximum on
a timescale of $T_{spike} \sim 6-8$ seconds, and then fade rapidly
away (Figure 6). During the spike, there are occasional episodes of
rapid large-amplitude fading in the count rates, such as seen at the
end of the spike shown in Figure 3 (Figure 7a).  The timescales for such
episodes are remarkably uniform. Whenever a drop of large enough
amplitude ($> {1 \over{e}}$) is observed, it has an e-folding
timescale of $\tau \sim 32$ ms (Figure 7b).  After examining $\sim 20$
such fading episodes on 14 August 1997 and on 9 September 1997, we
find a full range of $\sim 25-40$ ms for the e-folding timescales of
these events.  Based on the spectrosopic results above, the X-ray
spikes are found to have typical X-ray fluences of $E_X \simeq 4
\times 10^{39} \ {\rm ergs}$.

        The IR lightcurve (Figure 3b) shows large amplitude flares
that begin near the end of the X-ray hard dip \citep{Eiken98a}. They
are attributed to synchrotron emission from a plasma blob ejected at
relativistic speeds (\citet{FP97} ; \citet{Eiken98a}; \citet{Felix98}
; \citet{FP99} ; \citet{Vivek}), and similar to the larger ejections
seen by \citet{MR94}.  The IR lightcurve also shows a much fainter
excess associated with the oscillating portion of the X-ray lightcurve
\citep{Eiken00}.  The large IR flares reach their peak $\sim 1200$
seconds after the onset of the hard X-ray dip, or equivalently $\sim
600$ seconds after the X-ray spike.  \citet{Felix98} and \citet{FP99}
also both observed associated radio flares delayed $\sim 900$ seconds
from the IR flares.  Invoking a simple model for synchrotron emission
from an expanding plasma blob, \citet{Felix98} noted that the
extrapolated time for the onset of the blob is near the time of the
X-ray spike (see also \citet{Yadav}).

\section{The B-Bomb Scenario}

  We present here a magnetic bomb scenario associated with the
formation and disruption of an accretion torus, and its similarly
shaped magnetosphere, which accounts for the two time-scales of
\begin{eqnarray}
T_{dip}\simeq600\mbox{s},~~~ T_{spike}\simeq6\mbox{s}, 
\end{eqnarray}
the energetics, and the X-ray spectral evolution. This discussion is based 
on critical points of magnetic stability in the suspended accretion model 
for long gamma-ray bursts (\citet{mvp01}). We first present a cartoon 
outline of this scenario, and then identify specific features in the observed
lightcurves with physical events in the magnetic bomb scenario.

\subsection{Cartoon Outline}

We envision a rotating black hole and a thin accretion disk extending
to somewhere near the last stable orbit (Figure 8a).  As many have
contemplated, a seed magnetic field in the accretion disk may develop
a large-scale magnetic field by a dynamo process or, equivalently,
winding of magnetic field-lines. For instance, in the
magneto-rotational instability (MRI) of \citet{BalbusHawley},
differential rotation in a Keplerian disk will shear magnetic loops,
winding field lines and amplifying the magnetic field (Figure 8b).

   We conjecture that this leads to the creation of a net poloidal
magnetic flux, associated with an approximately uniform magnetization
of matter in the inner accretion disk. This produces a toroidal
magnetosphere around the black hole (Figure 8c). A torus magnetosphere
around the black hole with net poloidal flux is represented by {\em
two counter-oriented} current rings, concentric about the black hole
(Figure 1). Through this current configuration, most of the output of
a rapidly rotating black hole in equilibrium with the surrounding
torus magnetosphere satisfies 
(adapted from \citet{tho86,mvp99,mvp01})
\begin{eqnarray}
L_H\simeq  4\times 10^{39}\left(\frac{\eta}{0.1}\right) 
                      \left(\frac{B}{3\times 10^9\mbox{G}}\right)^2
                      \left(\frac{M_{BH}}{14M_\odot}\right)^2~\mbox{erg~s}^{-1}
\label{EQN_LH}
\end{eqnarray}
is incident on the inner face of the surrounding torus by the
equivalence to pulsar magnetospheres viewed in poloidal topology
\citep{gol69,mvp99}.  Here, $\eta=\Omega_T/\Omega_H$ denotes the ratio
of the angular velocity of the torus to that of the black hole, and
$B$ the strength of the poloidal component of the magnetic field.  The
MRI activity is initially driven by the power output of the black
hole, $L_H$.  The associated e-folding time for growth of the magnetic
field energy is about 0.2 s \citep{mvp03}, whereby the magnetic field
energy ${\cal E}_B$ rapidly builds up to equilibrium values on the
order of the kinetic energy ${\cal E}_k$ of the torus on the
time-scale of seconds. The magnetic field-energy ceases to grow in
this equilibrium state (see \citep{mvp03a} for a detailed calculation)
-- the dominant dissipation mechanism of the magnetic field is not
well-known, but probably is associated with turbulent reconnection as
considered in \citet{mvp01}.

In the weak magnetic field limit, a rotating black hole may provide a
torque to surrounding matter in the accreting state
\citep{ruf75}. Recalling that most of the black hole power output is,
in fact, incident on the inner face of a surrounding torus, we see
that this creates a ``magnetic wall" around the black hole
\citep{mvp03} blocking the passage of accreting material through the
inner disk.  This is the ``suspended accretion'' state \citep{mvp01a}.
Matter then builds up in the torus at a rate determined by the rate of
accretion in the extended disk (Figure 8d).  As the mass in the torus
increases, so does the associated equilibrium total magnetic
field-energy ${\cal E}_B$.  At the same time, the black hole gradually
spins down by conservation of energy and angular momentum. This
establishes causality in the extraction of black hole spin energy by
horizon Maxwell stresses as proposed in \citep{ruf75,bla77} -- see
also \citet{mvp03}.

  We here consider the possibility that this evolution reaches a
critical point of instability, which triggers the prompt release of
the stored magnetic energy ${\cal E}_B$ -- the ``magnetic bomb"
explosion (Figure 8e).  We equate the total energy released with the
energy ${\cal E}_X$ in the observed X-ray spike.  As individual field
lines disconnect from their anchor point, poloidal magnetic field
dissipates on a dynamical timescale. In this process, particles
accelerate, producing X-ray emission from synchrotron, curvature, and
inverse Compton processes. While individual field lines disappear
quickly, the overall disconnection and release of energy over the
entire region occurs in a sound crossing time. This is expected to be
accompanied by outgoing shock fronts in the accretion disk.  By
continuing accretion from the outer disk, the system gradually returns
to our inital state (Figure 8f), and the cycle begins anew.

\subsection{B-Bomb Features in the Lightcurves}

	Our cartoon highlights the following connections to the
observations of Type B events.

{\em The hard dips} $[T_{dip}\simeq 600$s$]$ represent the suspended
accretion state -- a uniformally magnetized torus pressing against a
magnetic wall around the black hole (Figure 8d).  Halting of accretion
through the hot inner disk results in a dip in the X-ray
lightcurve. The suspended accretion torus evolves on the long
timescale of accretion towards increasing mass and associated magnetic
field-strength. As shown in \citet{mvp03}, this magnetic field reaches
its equilibrium value on the time-scale of seconds when driven by
input from the black hole. While the equilibrium magnetic field
strength is not known ab initio, it is subject to a fundamental limit
set by the condition of magnetic stability of the torus against
buckling, given in terms of the Van Putten-Levinson criteria
\citep{mvp03}
\begin{eqnarray}
\left.\frac{{\cal E}_B}{{\cal E}_k}\right|_c < ~1/15
\label{EQN_EE}
\end{eqnarray}
of the poloidal magnetic field energy-to-kinetic energy.  (Note that
\citet{mvp03} also consider a tilt instability with a value of $1/12$.
Since these are similar values, we take the smaller limit here for
simplicity.) In what follows, we express the critical magnetic field
energy as
\begin{eqnarray}
\left. {\cal E}_B = \epsilon_B \ {\cal E}_k \right.
\end{eqnarray}
where $\epsilon_B < 1/15$ corresponds to magnetic stability.

{\em The X-ray spike} $[T_{spike}\simeq 6$s$]$ at the end of the dip
corresponds to the disconnection event and the associated sudden
release of the energy stored in the magnetic field (Figure 8e). This
is triggered by instability of the inner torus magnetosphere,
associated with violating the stability bound (\ref{EQN_EE}).  The
observed IR/radio synchrotron emission give considerable evidence for
prompt jet ejecta from this disconnection event.

{\em Recovery} begins promptly following the spike, as the inner disk
begins to refills the suspended accretion zone (Figure 8f). This
accounts for the observed X-ray recovery phase.

\section{Comparison with Observations}

	In the sections below, we discuss the detailed qualitative and
quantitative features of the Type B events in the context of the
magnetic bomb scenario, including spectral evolution, timescales, and
energetics.

\subsection{X-ray Spectral Evolution}

	The X-ray spectral evolution observed from GRS 1915+105 during
Type B events matches expectations for the magnetic bomb model. At the
beginning of the hard dip, the blackbody temperature at the inner
edge of the disk decreases as the disk inner radius itself expands
(Figure 4).  This behavior was first noted in GRS 1915+105 by
\citet{Tomasso1}.  The power-law emission component hardens at the
same time.  This is the initiation of the suspended accretion state,
during which the torus mass and its associated kinetic energy and
magnetic field-strength grow, increasing the energy stored in the
``bomb''. As the inner edge of the optically-thick accretion flow
moves to larger radii, the apparent inner radius of the disk is moved
out with an accompanying temperature drop, in accordance with the
observations.

The torus thus formed is connected to the extended accretion disk
through a shock front. The formation of this shock in the accretion
zone alters the non-thermal emission from the system, providing the
observed hardening.  The magnetic field in the inner region may also
provide some additional non-thermal X-ray emission from synchrotron
processes. The blackbody component quickly reaches an apparent
equilibrium (although at the low temperatures of mid-dip, RXTE is not
very sensitive to rather large changes in $T_{disk}$ and $R_{disk}$),
while the power-law index and normalization show some weak evolution.
The inner radius of the disk (the boundary with the suspended
accretion zone) is at $R \sim 150-200 \ {\rm km}$.  Note, however,
that the radii inferred from the RXTE spectral fits should be viewed
as approximations, with poorly known systematic correction factors.

As the inner torus magnetosphere becomes unstable, it disrupts and
produces significant spectral evolution, notably the X-ray spike --
representing the magnetic bomb explosion. At the onset of the spike,
the inner disk radius begins to shrink and the blackbody temperature
goes up (Figure 5).  At the same time, the power-law normalization
goes up, while the index remains constant at first.  Once the spike
reaches its peak and rapidly fades, the power-law index softens
rapidly, and the normalization seems to stop its increasing trend
momentarily, while the blackbody trends continue on.  This behavior
shows the infall of the suspended accretion zone boundary as the
magnetic field strength decreases during the explosion.  The power-law
evolution during the spike is dominated by the explosion itself, and
only after this fades away does the underlying ``soft'' component
reveal itself.  Note that after the spike has ended, the spectral
evolution is essentially over -- thus, the spike seems to reveal an
actual state change in the X-ray emission of GRS 1915+105, rather than
simply a momentary ``blip'' in the lightcurve.  This is confirmed
explicitly by the analyses of similar events by \citet{Simone}.

	Based on this correspondence, the magnetic bomb scenario
seems compatible with the rather complicated X-ray spectral evolution
of the Type B events.

\subsection{B-Bomb Energetics and Timescales}

	The identification of certain observed features with
counterparts in the magnetic bomb scenario also enables identification of
timescales, particularly of the hard dip itself.  We assume here that
the observed energy release in the X-ray spike is equal to the
magnetic energy stored in the bomb itself:
\begin{eqnarray}
{\cal E}_B={\cal E}_X,
\label{EQN_BX}
\end{eqnarray}
where
\begin{eqnarray}
{\cal E}_B={4 \over{3}} \pi R^3 \ (B_{pol}^2 / 8 \pi)f_B,~~~
{\cal E}_X \simeq 4 \times 10^{39} \ {\rm ergs}.  
\end{eqnarray}
Here, $f_B$ is a filling factor of order unity which describes the
volume of the inner torus magnetosphere as a fraction of $4\pi
R^3/3$. In what follows, we shall take $f_B=1$.  The critical magnetic
field-strength for stability of the torus satisfies (\ref{EQN_EE}),
where ${\cal E}_k={M_{ID}M_{BH}}/{2R}$. This gives 
\begin{eqnarray}
G M_{BH} M_{ID} / R = 2\epsilon_B^{-1}{\cal E}_X,
\end{eqnarray} 
where $M_{BH} = 14M_{\odot}$ is the mass of the black hole and $M_{ID}$ is the 
mass of the inner disk region in question. From the X-ray spectral fits, we see
an apparent $R \sim 150-200 \ {\rm km}$ during long dips.  Solving
for $M_{ID}$, we derive an estimate of
\begin{eqnarray}
M_{ID}\sim 8 \times 10^{20} \ {\rm g}
\left(\epsilon_B/0.085\right)^{-1}
\label{EQN_MID}
\end{eqnarray}  
This is the amount of accreted matter needed to contain the magnetic bomb
energy. We note that the mass (\ref{EQN_MID}) thus accumulated in a torus
against the magnetic wall around the black hole satisfies
\begin{eqnarray}
M_{ID}=\frac{4\pi}{3}\rho R^3=
  7.5\times 10^{20}\mbox{g}\left(\frac{\rho}{0.1\mbox{g/cm}^3}\right)
  \left(\frac{R}{6M}\right)^{3}
  \left(\frac{M}{14M_\odot}\right)^3
  \left(\frac{\epsilon_B}{0.0667}\right)^{-1},
\end{eqnarray}
corresponds to an average density of about 0.1 g/cm$^3$.

The torus builds up while matter accretes against the magnetic wall
around the black hole. We can estimate the accretion rate $\dot{M}$ 
from the X-ray luminosity during the accreting state, where
$L_X \simeq 0.15 \dot M c^2$ refers to a canonical value. 
Between the hard dips, GRS 1915+105 typically shows 
$L_X \simeq 1 \times 10^{39} \ {\rm ergs \ s^{-1}}$, giving
$\dot M \sim 6 \times 10^{18} \ {\rm g \ s^{-1}}$. Set by the outer 
accretion disk, the accretion rate is continuous. Hence,
$\tau_{acc} ={M_{ID}/{\dot M}}$ defines the time to build the torus
in the suspended accretion state, i.e.:
\begin{eqnarray}
\tau_{acc} \ga 130 \ {\rm s}
\left(\frac{\epsilon_B}{0.0667}\right)^{-1}.
\label{EQN_tau}
\end{eqnarray} 
This lower limit is in good agreement with the observed timescales
$\sim 500-700 s$ of hard dips.

As a cross-check on the energetics, we may rewrite (\ref{EQN_LH})
following the condition (\ref{EQN_BX}),
\begin{eqnarray}
L_H\simeq  3\times 10^{39}\left(\frac{\eta}{0.1}\right)^{3} 
                      \left(\frac{{\cal E}_X}{4\times10^{39}\mbox{erg}}\right)
                      \left(\frac{14M_\odot}{M}\right)~\mbox{erg~s}^{-1}
\label{EQN_LH2}
\end{eqnarray}
as an upper bound in the limit of rapid rotation of the black hole. In
the suspended accretion model, most of the black hole-luminosity $L_H$
is incident onto the inner face of the torus, for reprocessing in
magnetic winds and dissipation. Dissipation is due to turbulent
magnetohydrodynamical flow, in response to competing torques on the
inner and the outer face of the torus \citep{mvp99}.  These competing
torques promote super-Keplerian and sub-Keplerian motions, and hence
the torus assumes a geometrically thick shape. The modal power density
spectrum of magnetohydrodynamical turbulence will be flat up to the
first geometrical break in the azimuthal mode-number $m$, set by the
ratio of the major-to-minor radius of the torus $a/b$
\citep{mvp01,mvp03a}.  We may approximate the angular velocity
distribution in the torus by $\Omega(r)=\Omega_T(R/r)^q$, where the
rotation index $1.5<q<2$ is bounded between the Keplerian value and
Rayleigh's critical value for stability of the $m=0$ mode. In this
event, the balance equations describing the suspended accretion state
give rise to a dissipation rate
\begin{eqnarray}
P_d\simeq q\left(\frac{b}{R}\right)L_H
<0.27\left(\frac{b}{M}\right)\left(\frac{\eta}{0.1}\right)^{3/2}
\label{EQN_PD}
\end{eqnarray}
which is carried off in X-ray emissions, while the remainder of
$L_{wind} = L_H - P_d$ is carried off in magnetic winds.  (Here, $b$
denotes the minor radius of the torus in the equatorial plane).  These
winds may in fact be the root source of the observed jets associated
with the Type B events.

For GRS 1915+105 during the hard dips, we typically observe an X-ray
luminosity $L_X \simeq 3 \times 10^{38} \ {\rm ergs \ s^{-1}}$. Our
theoretical prediction of the upper bound $L_X\simeq 10^{39}$ erg
s$^{-1}$ according to (\ref{EQN_PD}) and (\ref{EQN_LH2}) is in good
agreement with this observational constraint for rapidly rotating
black holes, and nominal values $\eta\sim 0.1$ (corresponding to
$R\simeq 7M_\odot)$ and $b\simeq M$.

We also note that if the magnetic torus winds are baryon-rich (see
Figure 2), they will decrease the effective $\dot M$ during the
suspended accretion state.  According to (\ref{EQN_PD}), the observed
X-ray luminosity during the hard dip $L_X[sas]\simeq 3\times
10^{38}$erg s$^{-1}$ represents about 25\% of the combined luminosity
in X-rays and winds from the torus.  Hence, we estimate $L_{wind}
\simeq 1\times 10^{39}$erg s$^{-1}$ for the torus wind.  This may
carry off matter at an approximate rate of $\dot{M}_{wind}={L_{wind}
\over {c^2}}$, or $\dot{M}_{wind}=3.3\times 10^{18}$g s$^{-1}$.  This
would then revise our estimate of the time required to build the torus
(\ref{EQN_tau}) to:
\begin{eqnarray}
\tau_{acc} > 300 \ {\rm s}
\left(\frac{\epsilon_B}{0.0667}\right)^{-1}.
\end{eqnarray} 
This lower bound is again in agreement with the observed timescales
$\sim 500-700 s$ of hard dips.

 \subsection{Timescales for the X-ray Spike}

       The spike itself shows two important timescales which can be
explained under the magnetic bomb scenario.  As noted above, the X-ray
spike emission is identified with the disruption of the inner torus
magnetosphere surrounding the black hole, releasing its magnetic
energy on the dynamical timescale of the torus.  The field lines
should not all disconnect simultaneously, but rather over the sound
speed crossing time of the inner disk region.  That is, a single field
line disconnects, creates a magnetosonic wave disturbance that
propagates through the disk causing subsequent field lines to
disconnect, and initiates a chain reaction propagating at the sound
speed. For the limited ratio of magnetic-to-kinetic energy of the
flows under consideration here, the magnetosonic speed is close to the
sound speed. For a disk temperature of $\sim 0.5$ keV, the
corresponding sound speed is $c_s \sim 100$ km/s.  Thus, the sound
crossing time over a diameter of $\sim 400$ km is $t_s \sim 4$
seconds.  Given the crude estimation we are making, this is a very
close match to the $\sim 6$s duration of the X-ray spikes.

	As the disconnection wave progresses, it encounters
inhomogeneities in the B-field.  Whenever there is an absence of field
lines to detach and radiate, the observed X-ray flux should fade on
the timescale of the radiative lifetime (approximately equal to the
dynamical timescale) of the last detached field lines.  For a radius
of $\sim 200$ km, the corresponding dynamical timescale for a black
hole of $14 M_{\odot}$ is $\sim 15$ ms.  The observed range in
timescales for fading events during the spike is $\tau_s = 25-40$ ms,
corresponding to Keplerian radii of $R = 300-400$ km.  Again, for such
crude estimation this is a very close match.

\section{Discussion}

	This magnetic bomb scenario for GRS 1915+105, based on the
more detailed suspended accretion model for gamma-ray bursts from
rotating black holes \citep{mvp01}, seems to provide a good match to
many features in multiwavelength observations of Type B jet events.
To our knowledge, it is the first such physical model to do so at the
level of qualitative and quantitative lightcurve/timing features,
spectral evolution, and the observed recurring timescales in the dips
and spikes.  Therefore, this physical framework seems to be a
promising one within which to proceed in investigating the formation
of relativistic jets in this system.  However, there is still
significant work to be done on this general scenario. Particular
outstanding issues in the work mentioned above include the problem of
creating net poloidal magnetic flux in forced shear flow and the
radiative physics of prompt disconnection events of magnetospheres.

Furthermore, there are complex observational phenomena associated with
the Type B events which we have not addressed here.  For instance,
\citet{Markwardt} showed that 1-10 Hz quasi-periodic oscillations
(QPO) are intimately linked to the system evolution during the hard
dip (suspended accretion state in our model), and disappear outside
the dip.  This leads us to advance the notion that Type B events
represent a magnetic flip-flop between a 'dirty' accreting state,
corresponding to the observed oscillatory/fluctuating state, and and a
'clean' suspended accretion state, corresponding to the observed
long-duration and smooth hard-dip.  We describe these states as
follows:

   (1) The dirty state is produced by an accreting inner disk of small
   radius with negligible or disordered poloidal magnetic field. The inner disk
   is turbulent and accretes on a viscous timescale. Only short-lived QPOs
   are allowed, by limited or short-range coherence in hydrodynamical
   wave-modes in the disk, whose convective motion spirals into the black
   hole.

   (2) The clean state is produced by the suspended accretion state
   with a torus of extended radius and significant ordered poloidal
   magnetic field of about $10^9 G$.  The torus develops on the
   viscous timescale of accretion, while it supports an inner
   torus magnetosphere of increasing magnetic energy.  Long-lived QPOs
   are allowed (such as those of \citet{Markwardt}), by large-scale
   coherence in wave-modes in the torus, whose convective motion forms
   closed circular orbits.

   We attribute the transition from the accreting state to the
   suspended accretion state to a switch-on of poloidal magnetic
   flux. This transition is smooth and non-explosive. The timescale
   for pushing out the inner disk into the resulting torus is on the
   order of seconds, set by the black hole luminosity onto the inner
   disk.  We have not determined the cause of the appearance of such
   ordered magnetic field, or the trigger thereof. We attribute the
   transition from the suspended accretion state back to the ``dirty''
   accreting state to a switch-off of the poloidal magnetic flux: our
   B-bomb scenario. The timescale for remnants of the torus to fall
   back and form an inner disk (of smaller radius) is a viscous
   timescale, possibly shortened by additional angular momentum loss
   in magnetic winds.

   Other observational phenomena not treated here during these events
include the fact that Type B infrared flares often show significant
substructure ( see e.g. \citet{Eiken98a}), spectral line evolution
\citep{Eiken98b}, as well as extended low-level excesses
\citep{Eiken00}.  Given the link between the disconnection event
(X-ray spike) and the launching of the long-wavelength flares under
our scenario, we also expect that there may be some correlation
between these IR flare properties and features in the X-ray spike.

	We also mention here a conceivable link between the detailed
observations and other models detailing various behaviors in GRS
1915+105 and microquasars in general.  For instance, the Accretion
Ejection Instability (AEI) model \citep{Tagger} shows some promise for
explaining the 1-10 Hz QPO above in terms of a spiral shock wave in
the disk (\citet{JRodriguez} ; \citet{Varniere}).  Conceivably, our
magnetic bomb scenario is compatible with a ``magnetic flood''
suggested in the context of AEI by \citet{Tagger2}.  Furthermore, our
scenario explains invokes a strong poloidal field component in the
inner disk.  Recent MHD models of relativistic jet formation in the
innermost radii around a Kerr black hole consider a similar agent to
produce powerful jets such as those observed in GRS 1915+105
\citep{Meier}.

Finally, we point out that the timescale arguments above may be
applicable to bomb models in general.  Specifically, the X-ray spike
duration ($\sim 6$s) corresponds to the sound crossing time for the
inner disk region defined by the X-ray spectral fits ($R \sim 200$
km).  In addition, the rapid fading timescale ($\sim 32$ms) observed
during the spikes is similar to the dynamical timescale at this same
radius.  Finally, the energy released in the spike is equivalent to
$\sim 10 \%$ of the accretion energy accumulated during the duration
of the hard X-ray dip immediately preceding it.

\section{Conclusions}

	We have presented a magnetic bomb scenario for Type B
relativistic jet events in GRS 1915+105, based on the more detailed
suspended accretion state model of \citet{mvp01} originally developed
for gamma-ray bursts.  This scenario posits episodic formation and
disruption of a torus and a similarly shaped magnetosphere in response
to the interaction between a rotating black hole and the inner
accretion disk.

The inner disk couples to the black hole, receiving energy and angular
momentum from it.  This results in a suspended accretion state,
halting accretion through the (X-ray-emitting) innermost disk radii,
and storing magnetic energy in the resulting torus.  This is observed
as the $\sim 600$s hard X-ray dip.  Once this energy level reaches a
critical value, the inner disk and black hole disconnect through a
magnetic buckling instability, and the stored magnetic energy is
released -- the ``B-bomb'' explodes.  We identify the X-ray spike with
the prompt disconnection of the inner torus magnetosphere.  After the
explosion, the inner disk refills and the cycle begins anew.  This
scenario provides general agreement with three observed timescales,
energetics and spectral evolution in the X-ray waveband, as well as
with key features in infrared and radio wavebands.

\acknowledgements SSE is supported by an NSF CAREER grant
(NSF-9983830), SSE thanks D. Rothstein for discussions and suggestions,
and MVP thanks A. Levinson for discussions. The LIGO Observatories
were constructed by Caltech and MIT with funding under cooperative
agreement PHY 9210038. The LIGO Laboratory operates under cooperative
agreement PHY-0107417. This paper has been assigned LIGO document
number LIGO-P03xxxx-00-R.

\vfill \eject

\begin{figure}
\plotone{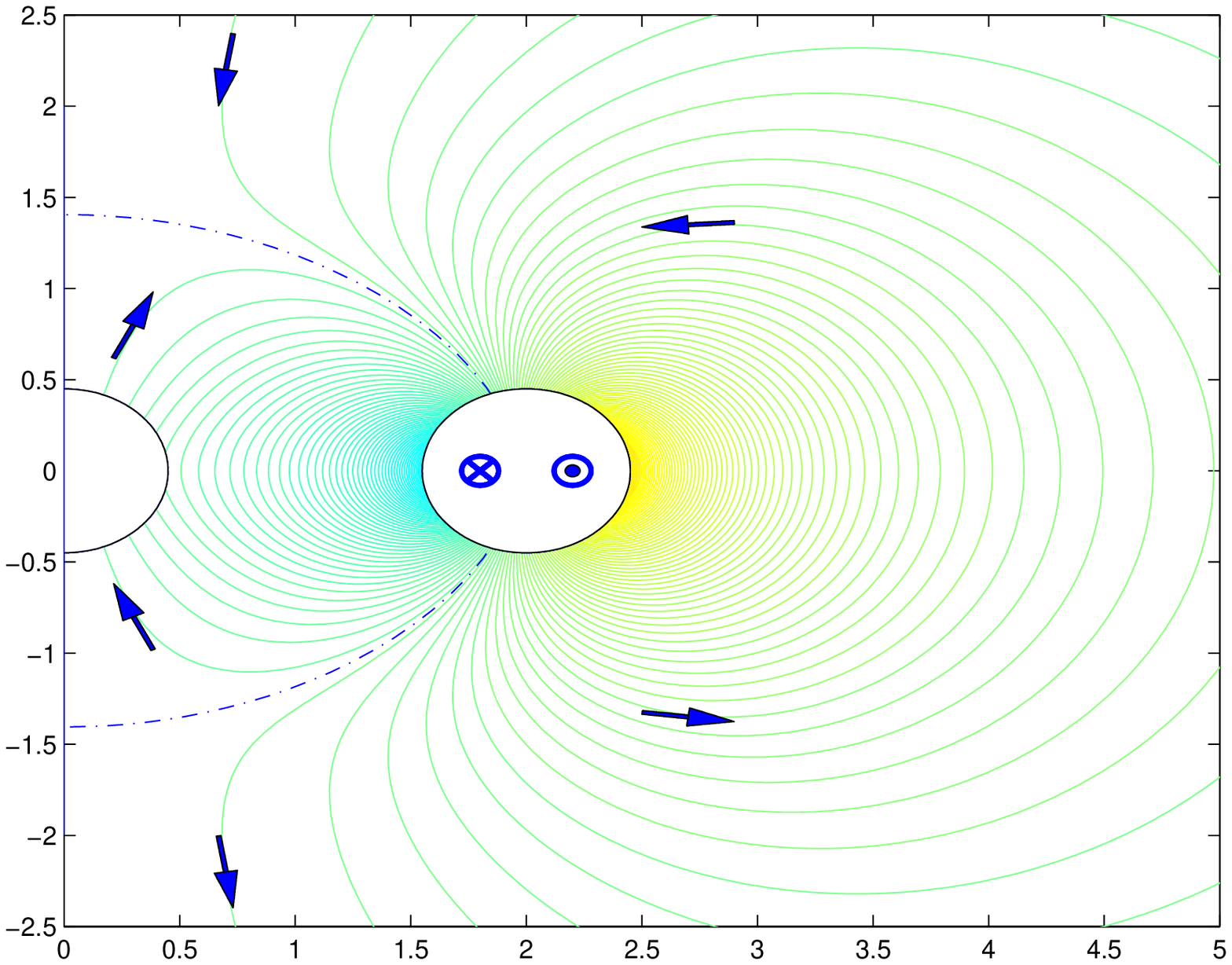}
\caption{Numerical solution of the poloidal topology of magnetic flux-surfaces in 
vacuo and flat spacetime, 
produced by two concentric and oppositely oriented current rings in the torus
({\em center}) and a central current ring representing the equilibrium magnetic moment 
of the central object ({\em left}). The dashed line is the separatrix between 
the inner and the outer torus magnetospere, associated with the inner and outer faces
of the torus. A Kerr black hole develops a magnetic moment which 
preserves essentially uniform and maximal magnetic flux through its horizon in
equilibrium with the surrounding torus magnetosphere.
Torus winds, when sufficiently powerful, may alter the poloidal topology by 
creating a tube of open field-lines to infinity for baryon poor outflows
from the black hole. (Reprinted from M.H.P.M. van Putten
\& A. Levinson, {\em The Astrophysical Journal}, 584, 937 (2003)).}
\end{figure}

\begin{figure}
\plotone{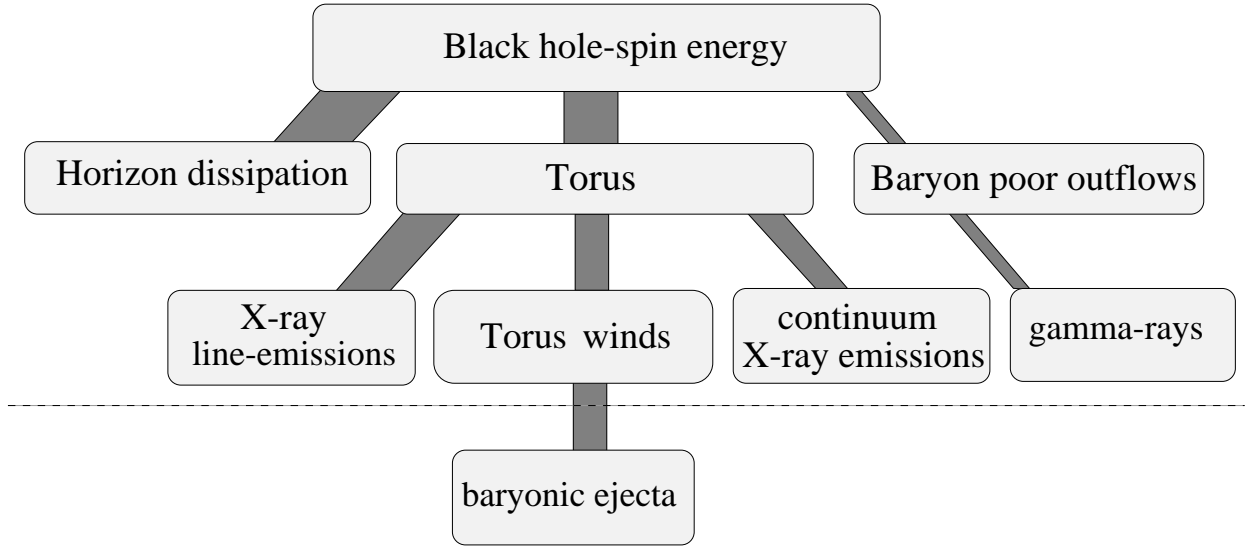}
\caption{A radiation-energy diagram for the interaction of a Kerr
black hole with a surrounding uniformly magnetized torus applied to
GRS 1915+105, based on the suspended accretion model for GRBs from
rotating black holes \citep{mvp01}. The ratio of magnetic field
energy-to-kinetic energy in the torus is subject to the van
Putten-Levinson stability criterion ${\cal E}_B/{\cal E}_k<1/15$
\citep{mvp03}.  Most of the rotational energy is dissipated in the
event horizon of the black hole, and this dissipation sets the
lifetime of rapid spin of the black hole. Most of the black hole
luminosity is incident onto the inner face of the torus for
reprocessing into various channels, including thermal emissions,
line-emissions \citep{rey03}, and build-up of poloidal magnetic
field. The fate of torus winds is uncertain or unkown (below the
dashed line), and may be related to baryonic ejecta and collimating
winds.  In the proposed B-bomb scenario, we identify observed energy
release ${\cal E}_X$ in soft X-rays in the short duration X-ray spike
of Type B events with the dissipation of magnetic field-energy ${\cal
E}_B$ following prompt disconnection from the central black hole.  The
active lifetime of of GRS 1915+105 for this behavior is actually set
by the mass-donating star, since the magnetic field energy permits a
lifetime of tens of Gyr for the rapid spin of the black hole.}
\end{figure}

\begin{figure}
\plotone{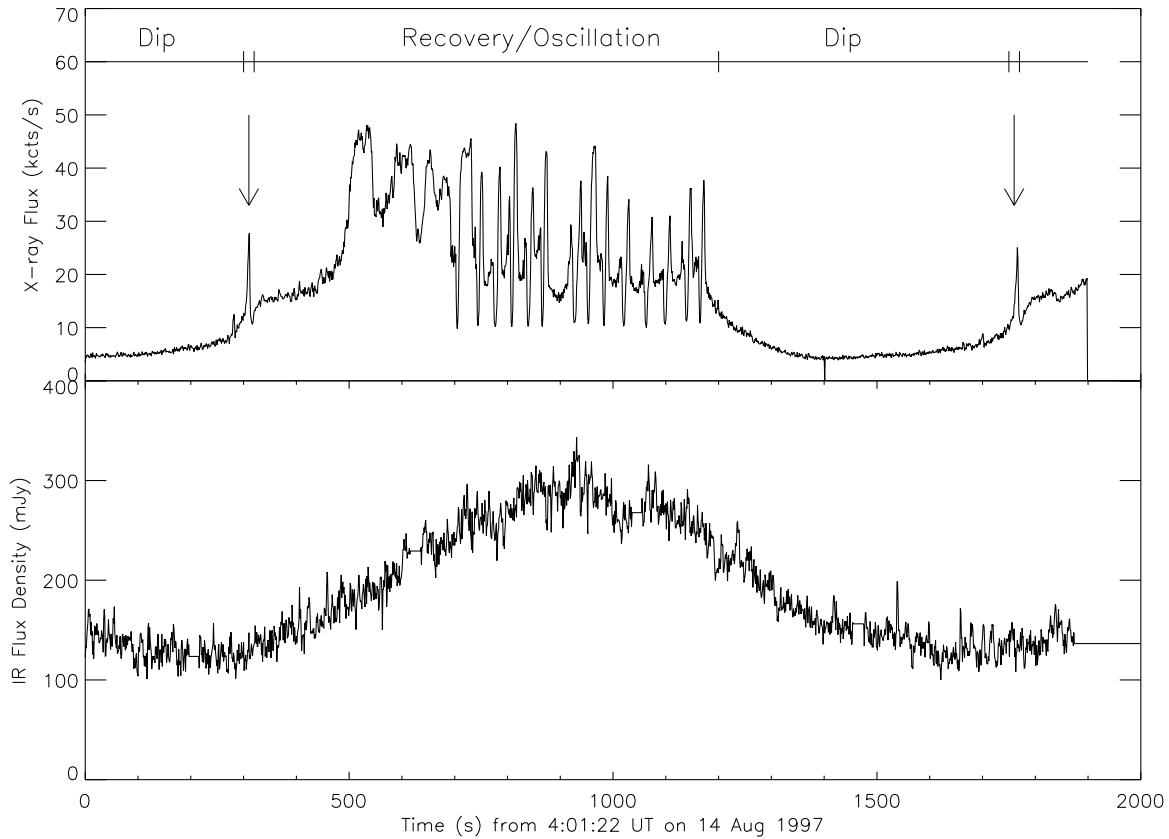}
\caption{X-ray and IR observations of a Type B event based on the
observations of \citet{Eiken98a}.  The timeline and labels at top show
the various states described in the text, while arrows indicate the
X-ray spikes.  The typical Type B pattern begins with the onset of a
spectrally-hard X-ray dip, where the emission is dominated by a
non-thermal power-law spectral component.  Near the end of the dip,
there is a spectrally-soft X-ray spike, followed by a ``recovery'' ($t
= 300-700$s above) which is dominated by thermal emission from the
accretion disk (see e.g. \citet{Tomasso1}).  The X-ray lightcurve then
transitions into an oscillating state ($t=700-1200$s) characterized by
rapid changes in the blackbody spectral component coupled with soft
flares in the power-law component \citep{Eiken00}.  Finally, a new
hard dip begins ($t=1200-1900$s).}
\end{figure}

\begin{figure}
\plotone{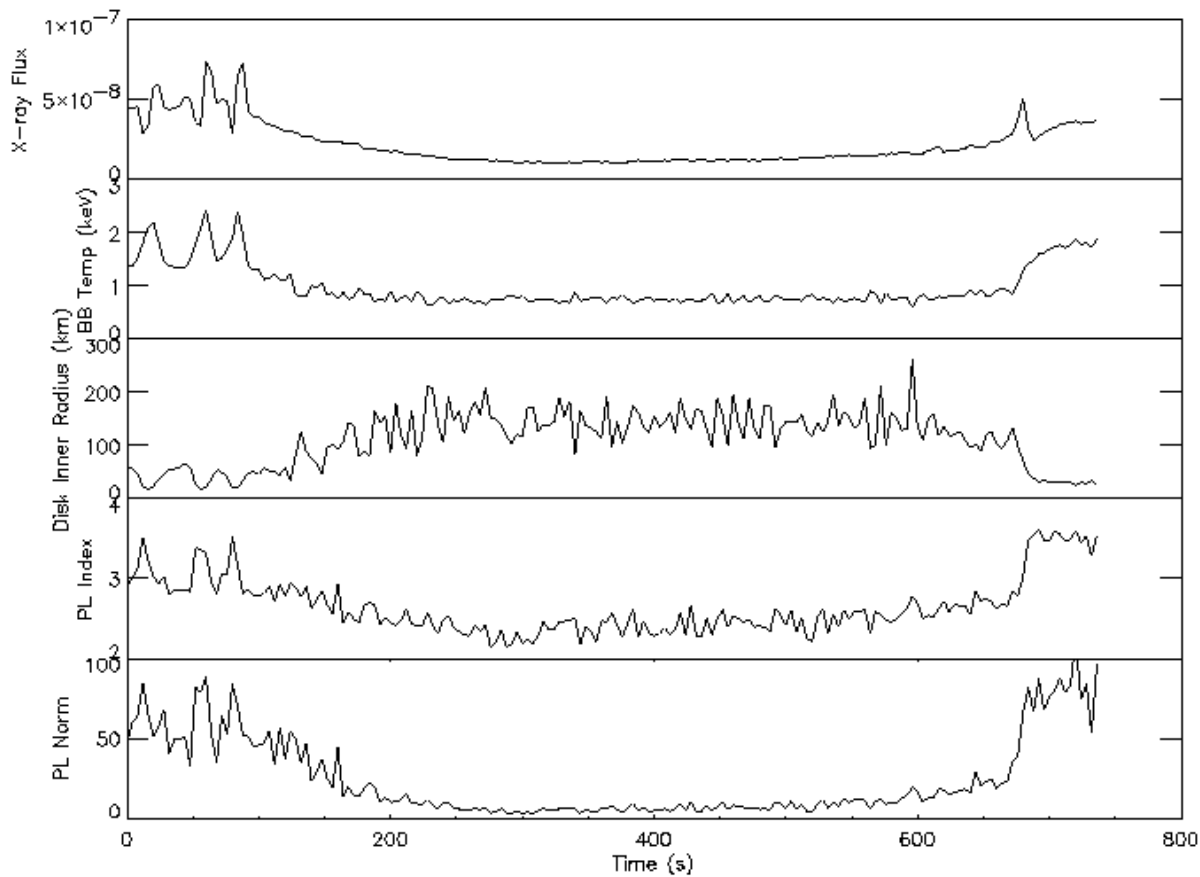}
\caption{X-ray spectral evolution during a Type B event hard dip.
Spectral fits are performed at 1-s time resolution to RXTE PCA data as
described in \citet{Eiken98a}.  At the beginning of the dip, the inner
accretion disk blackbody emission changes, with the temperature at the
inner edge dropping while the radius of the inner edge increases
dramatically.  The power-law component hardens while its normalization
drops.  The dip terminates suddenly at the X-ray spike.  Note that
while the spike appears to have a short duration in the light curve,
it actually reveals a dramatic ``one-way'' state change in the X-ray
spectral properties.}
\end{figure}

\begin{figure}
\plotone{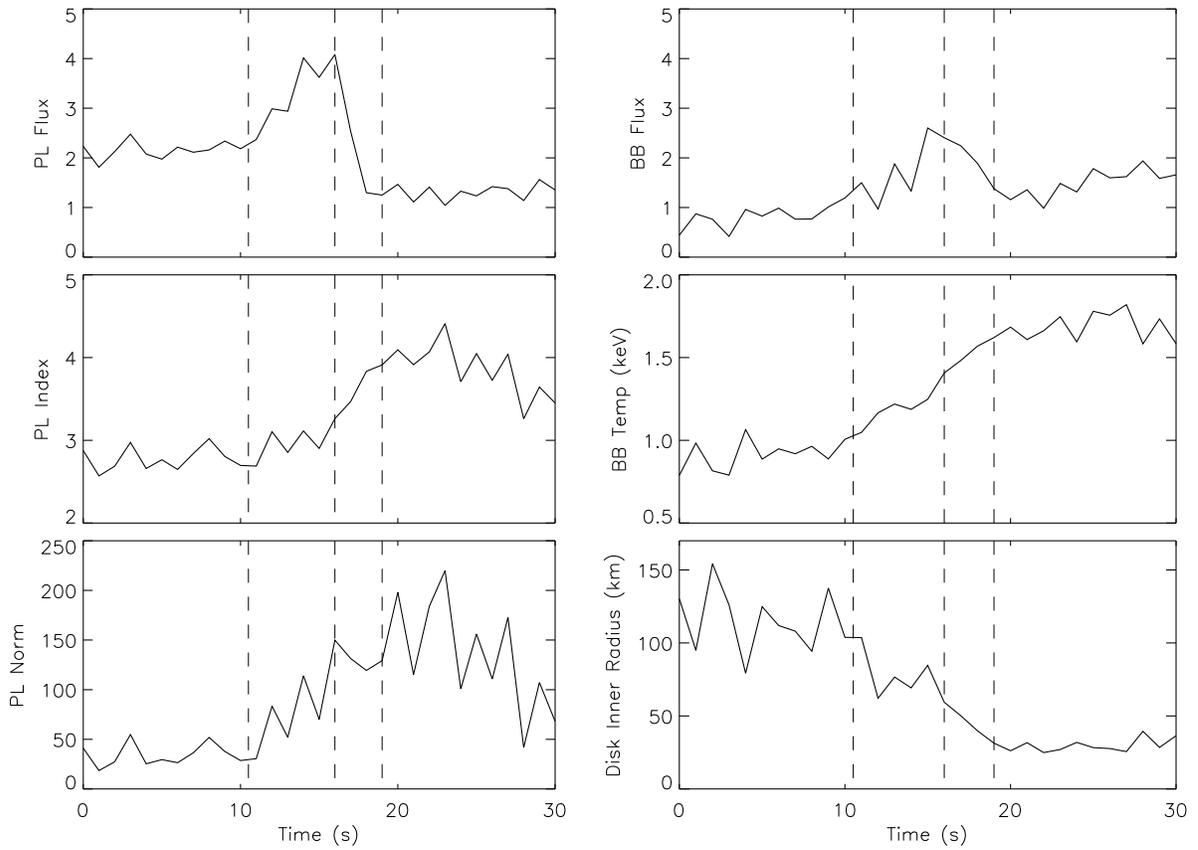}
\caption{X-ray spectral evolution during an X-ray spike with 1-s time
resolution.  The first time tick indicates the beginning of the spike.
The second the peak of the spike, and the third the end of the spike.}
\end{figure}

\begin{figure}
\plotone{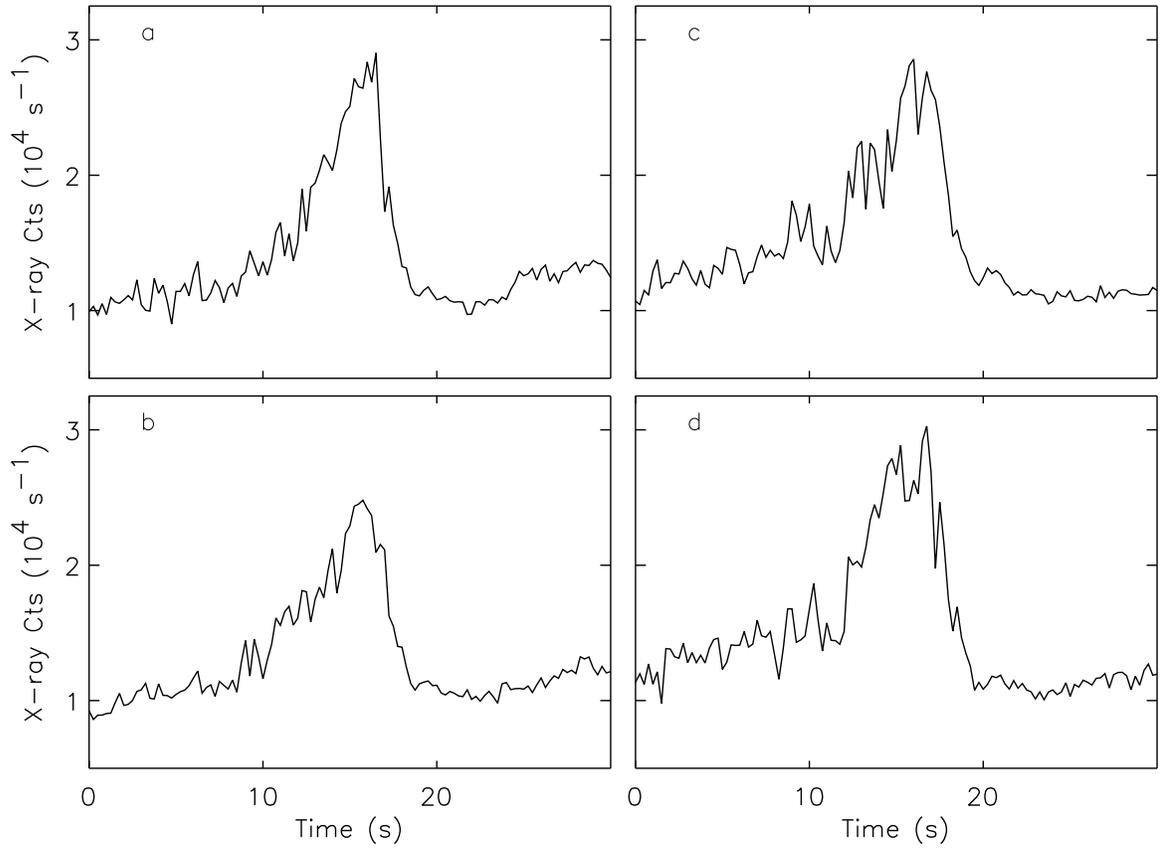}
\caption{\it Lightcurves of 4 X-ray spikes obsered with the RXTE PCA
on 14 August 1997.  Note that the typical spike duration is $\sim
6-8s$ for essentially all of these spikes.  In addition, there is
statistically significant rapid ($<1$s) variability during the spikes
themselves.}
\end{figure}

\begin{figure}
\plotone{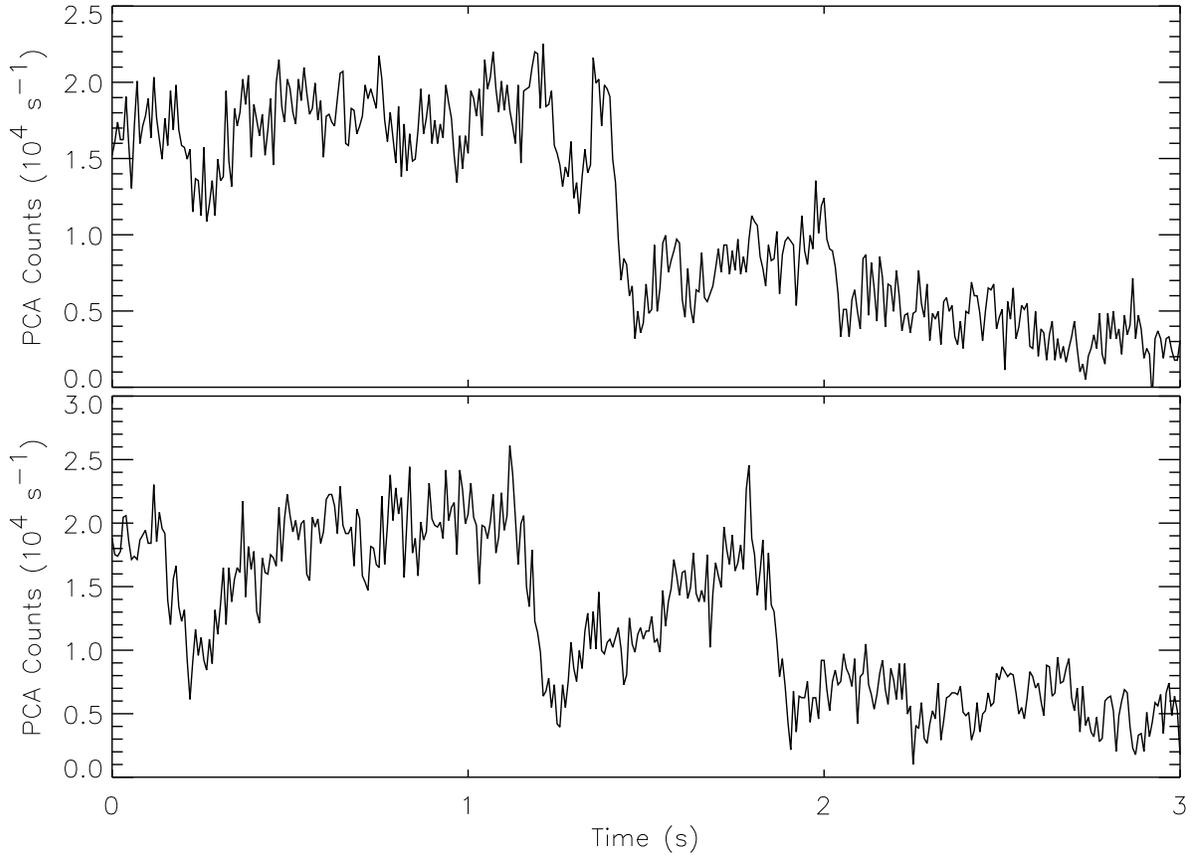}
\caption{\it High time-resolution (8-ms bins) PCA lightcurves of X-ray
spikes from (a) Figure 4a (b) Figure 4d.  Note the large-amplitude
rapid fading events indicated by the arrows.  All of the events with
large fractional amplitudes ($\geq e$) and monotonic drops have
e-folding timescales of $\tau \simeq 32$ ms.}
\end{figure}

\begin{figure}
\plotone{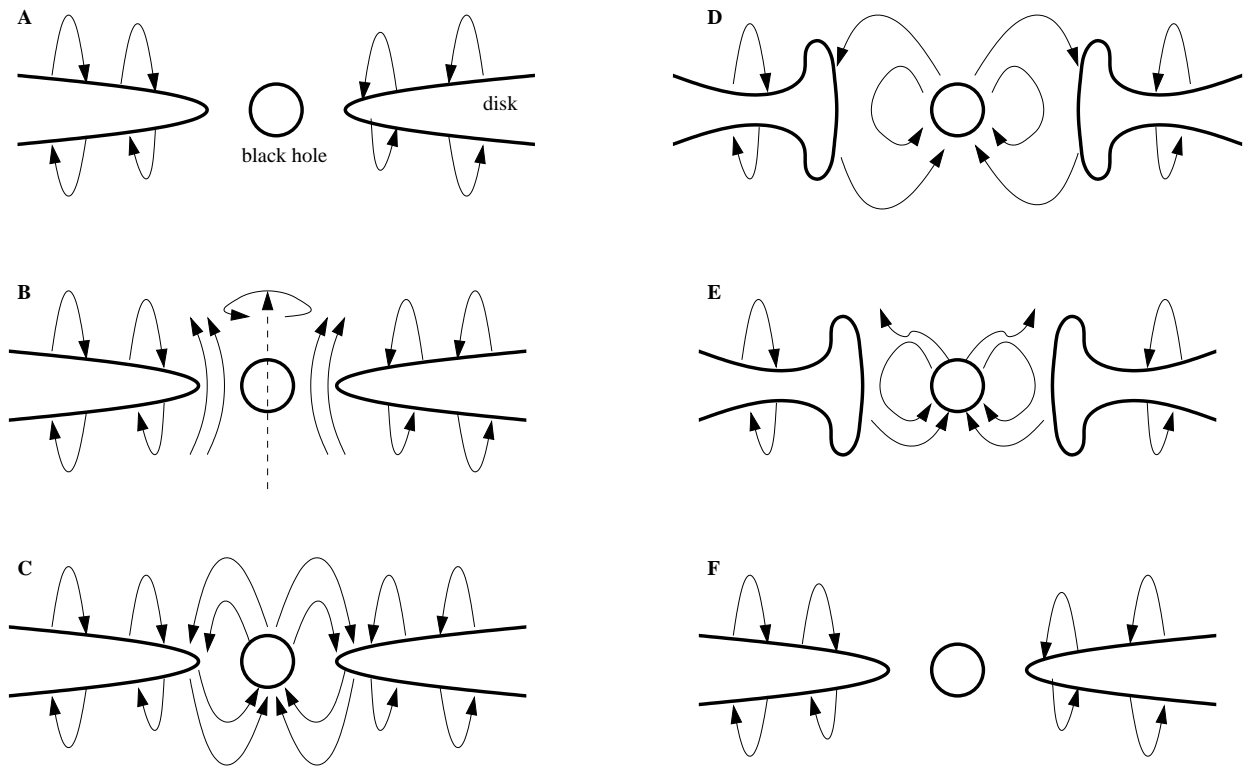}
\caption{Cartoon of magnetic bomb evolution.  The system begins with a
thin accretion disk extending down to near the last stable orbit of a
rotating black hole (A).  Through a magneto-rotational instability, an
ordered magnetic field develops and grows (B).  Eventually, this field
causes a connection between the inner accretion disk and the rotating
black hole (C).  This connection transfers energy and angular momentum
from the black hole to the inner disk, suspending the accretion of
material through the inner disk and producing a torus of material (D).
When the ratio of magnetic energy to kinetic energy reaches the van
Putten-Levinson instability criterion, the field disconnects and the
stored magnetic energy is rapidly disippated (the ``B-Bom'' -- E).
The inner disk then refills on a viscous timescale, and the system
return to the initial state, allowing the cycle to begin anew (F).}
\end{figure}

\end{document}